\documentstyle{mn}
\begin{document}
\title
[HST imaging and ground-based spectroscopy of nova shells -- I] 
{Hubble Space Telescope
imaging and ground-based spectroscopy of old nova shells -- I. 
FH Ser, V533 Her, BT Mon, DK Lac, V476 Cyg}
\author[C.D.\ Gill \& T.J.\ O'Brien]
{C.D.\ Gill$^1$ \& T.J.\ O'Brien$^2$ 
\\
1. Astrophysics Research Institute, Liverpool John Moores University, 
Twelve Quays House, Egerton Wharf, Birkenhead, L41 1LD\\
2. Jodrell Bank Observatory, The University of Manchester, Macclesfield, Cheshire, SK11 9DL
}
\maketitle
\begin{abstract}
In this paper we report on the first five out of eleven observations in
our programme of Hubble Space Telescope (HST) imaging of old nova shells.
We present new WFPC2 images of the shells around FH Ser and V533 Her 
taken in the F656N
(H$\alpha$+[N~II]) filter. We also show long-slit spectra taken using
the William Herschel Telescope (WHT) in La Palma for these objects in 
the same spectral range.

The shell around FH Ser is found to be a prolate ellipsoid of
ellipticity 1.3$\pm$0.1 inclined at 62$\pm$4$^\circ$ to the line of
sight.  The shell has an equatorial ring which is found to be due to
increased emission in the two [N~II] lines rather than H$\alpha$. The
expansion velocity is best modelled by a true equatorial expansion
rate of 490$\pm$20~km~s$^{-1}$. The best-fitting systemic velocity is
$-$45~km~s$^{-1}$.  A synthetic image and synthetic spectra are also
presented for this model for comparison with our observations. We
derive a distance to FH Ser of 950$\pm$50 pc. The origin of the [N~II]
equatorial ring is discussed in the context of a photoionization
feature resulting from aspherical illumination by the central source
rather than a simple density enhancement. It is possible however that
the ring is also in part due to an extremely localised increase in the
nitrogen abundance. The brightest part of the shell was found to have
a surface brightness of $9.1\times
10^{-15}$~erg~cm$^{-2}$~s$^{-1}$~arcsec$^{-2}$.

Similar imaging and spectroscopy of the nova V533 Her reveal a shell of 
radius $5\pm0.7$~arcsec with an axial ratio of $1.2\pm0.2$ and peak 
surface brightness
$1.3\times 10^{-15}$~erg~cm$^{-2}$~s$^{-1}$~arcsec$^{-2}$. The expansion
velocity of this shell is $850\pm150$~km~s$^{-1}$ and the distance is estimated
to be $1250\pm300$~pc. 

The shells around BT Mon, DK Lac and V476 Cyg 
were not detected with HST implying 3$\sigma$
upper limits to the surface brightness in H$\alpha$+[N II] of 5.3--6.3$\times 
10^{-16}$~erg~cm$^{-2}$~s$^{-1}$~arcsec$^{-2}$.

\end{abstract}

\begin{keywords}
circumstellar matter -- novae, cataclysmic variables
\end{keywords}

\section {Introduction}

Every classical nova outburst should result in the ejection of
$10^{-5}$ to $10^{-4}\,M_{\odot}$ of material at velocities of the
order of hundreds to thousands of kilometres per second (Bode \& Evans
1989).  Study of these expanding shells of nova ejecta has importance
for a range of astrophysical areas including the physics of
thermonuclear reactions, mixing mechanisms, nebular shaping,
radiation-driven winds, clumping mechanisms, astrophysical chemistry
and the formation of dust. Novae
provide a real-time laboratory in which these processes can be
investigated. The work presented in this series of papers builds on our
ground-based imaging survey of the ejecta of old novae using 4-m class
telescopes (Slavin, O'Brien \& Dunlop 1995, Gill \& O'Brien 1998) in
two ways. Firstly we present high-resolution imaging obtained with the
Wide Field Planetary Camera 2 (WFPC2) on the Hubble Space Telescope
(HST). This is then complemented by kinematical information obtained from
spatially-resolved spectroscopy using the William Herschel and
Anglo-Australian Telescopes (WHT \& AAT respectively).

This first paper presents the HST images and medium-resolution long-slit
WHT spectra of the shells of FH Ser and V533 Her in the region of
H$\alpha$. We also present non-detections in the HST imaging of BT Mon,
DK Lac and V476 Cyg. 
\S \ref{obs} describes the observations and \S \ref{res}
presents the results. The data are interpreted and discussed in
\S \ref{int}. The corresponding observations of HR Del, RR Pic, T Aur,
V1500 Cyg and V842 Cen will be presented in papers to follow.

\section{Observations}
\label{obs}

Hubble Space Telescope (HST) observations of the novae were made during 
1997/98 -- see Table 1. In each case the nova was positioned 
on the Planetary Camera chip of WFPC2. The pixel scale is 46~mas with a complex
point spread function distributing 60\% of the flux from a point source over
the 3$\times$3 pixel region centred on the source.

The standard pipeline data processing was used (Leitherer 1995) and the
images combined eliminating cosmic rays and bad pixels using standard 
tasks in STSDAS/IRAF. Narrow band photometry was performed using the
prescription described by Dudziak and Walsh (1997).
For the images where there was no immediately obvious detection of an ejected
shell we applied a 7$\times$7 sliding-box median-filter to the images
to reduce the effects of noise and contamination from low level cosmic
rays.  This process enhanced the slower varying underlying background
which is more consistent with an extended nova shell and thus allowed
visual detection of shells to be investigated.

Long-slit spectra of the shells of FH Ser and V533 Her 
were taken on the 4.2m WHT in La Palma,
Canary Islands using the ISIS spectrometer on 1996 Aug 3 -- see Table 1. 
This instrument has the advantage of having both a blue and red arm which
allows simultaneous observations of two wavelength ranges with a suitable
dichroic. Using the TEK-1 and TEK-2 chips with the R1200B and R1200R
gratings this instrument has a spectral range of 420\AA~ and a dispersion
of 0.41\AA/pixel in both arms. Observations were taken with central
wavelengths of $\lambda \lambda$ 4950 and 6600\AA~ for the blue and red
arms respectively to allow primarily the lines of H$\alpha$ and H$\beta$
($\lambda \lambda$ 6562.8 and 4861.3\AA~respectively) to be examined as
well as [O~III] ($\lambda \lambda$ 4959 and
5007\AA), [N~II] ($\lambda \lambda$ 6549 and 6583\AA) and  He I ($\lambda
\lambda$ 4921, 5015 and 6677\AA). The dispersion along the slit was 0.36
arcsec per pixel. Slits of width 1~arcsec 
were used at position angles chosen to investigate features
revealed in previous imaging. The spectra were reduced using 
standard techniques.

\section{Results}
\label{res}
\subsection{FH Ser}
FH Ser (Nova Ser 1970) erupted in 1970 and was discovered by Honda
(1970). It reached a peak visual magnitude of 4.4 on Feb 18 (Burkhead et al. 
1970, Borra \& Andersen 1970).
The lightcurve shows a significant 
DQ-Her-type dip (e.g. Fig. 2 of Rosino et al. 1986) and is characterized by a 
$t_3$ time of 62 days (Duerbeck 1987). 
Recently it has been classifed as a `slow' nova belonging to the 
{\it Fe II} spectroscopic class (Della Valle \& Livio 1998, 
Williams 1992).
Duerbeck (1992) produced the first image
of the shell from which he measured a radius of 2.7 arcsec and
described an equatorial band of enhanced emission. He derived an
orbital inclination of 58$^\circ$ and a distance of 850 pc. Later
images were taken by Slavin at al.\,(1995) in 1993 September. Their
images were deeper and in several narrow wavebands covering
approximately $\lambda \lambda$ 6530 to 6608\AA . They also resolved
the equatorial enhancement and derived an inclination angle of
$58\pm14^\circ$. They suggested that the equatorial band is due to
H$\alpha$ emission from a density enhancement.

Della Valle et al. (1997) imaged this nova shell in 1996 March with a
H$\alpha$+[N~II] filter. They showed the shell to be elliptical with
outer diameters of 7.6 and 9 arcsec with a peak-to-peak minor axis of 5.4
arcsec. No peak-to-peak major axis was given. They also showed that the
shell had an equatorial ring. They derived an inclination angle of
60$^\circ$ and a distance of 870$\pm$90 pc.

\subsubsection{Imaging}
\begin{figure*}
\vspace {12.0truecm}
\includegraphics{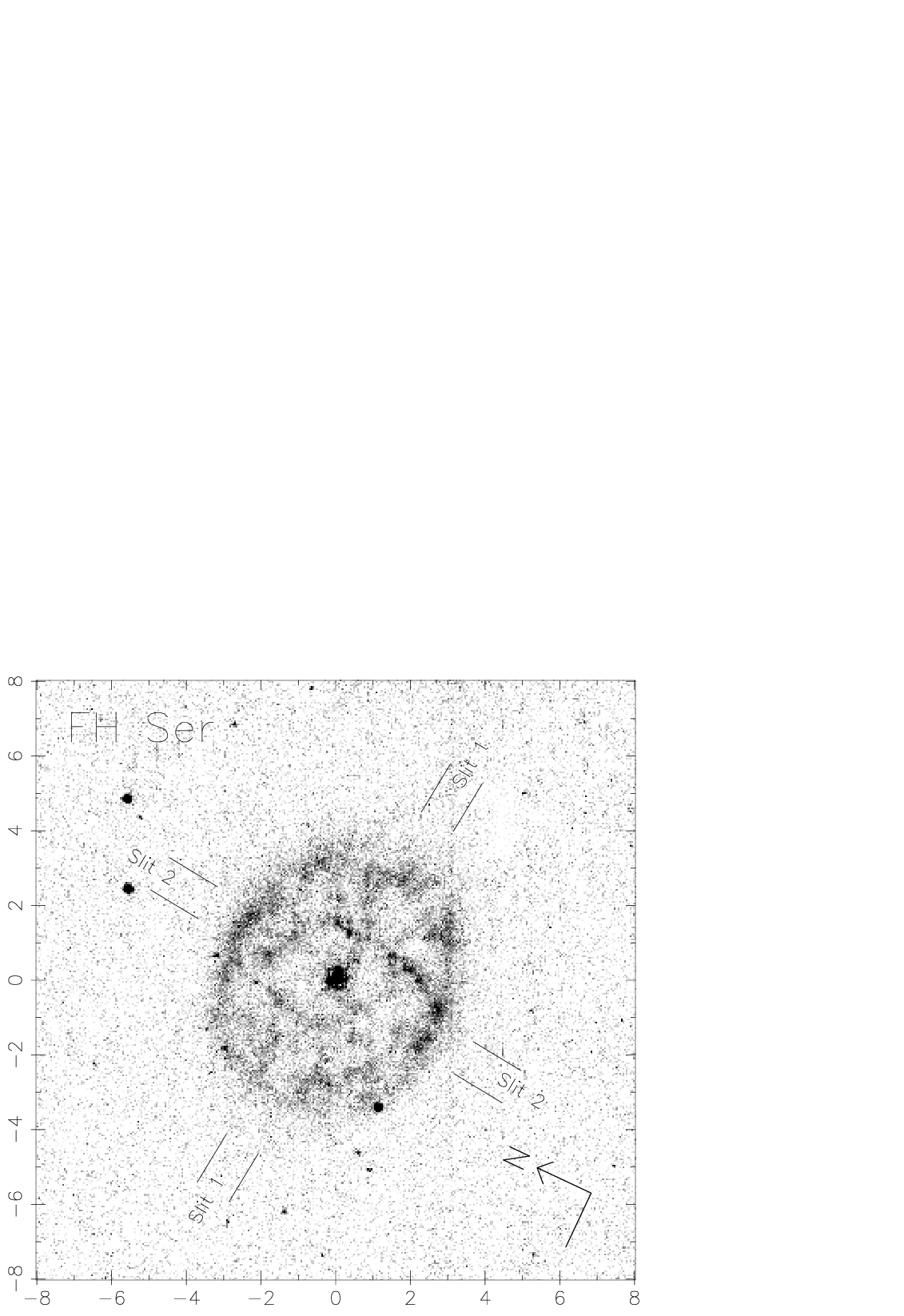}
\caption[]{HST F656N image of the remnant around FH Ser. The orientation
of the image is shown with the arrow towards north and the plain line to
east. The slit positions for spectroscopy are shown with 1 arcsec slits.}
\label{fh_image}
\end {figure*}

The FH Ser remnant was imaged with HST on 1997 May 11 with two 1200s
integrations, Fig.~\ref{fh_image}.  The maximum pixel value
attributable to the shell is 10.9 counts (a 7-$\sigma$ detection 
corresponding to a surface
brightness of 9.1$\times
10^{-15}$~erg~cm$^{-2}$~s$^{-1}$~arcsec$^{-2}$).
The shell takes the form
of a very clumpy limb-brightened prolate ellipsoid. There is a clear
equatorial ring which is brighter in the western half compared to the
eastern. If we assume this equatorial enhancement is an inclined
circular ring then the derived inclination angle is
62$\pm$1$^\circ$. The apparent major axis is 7.0$\pm$0.3 arcsec and
the minor axis 5.8$\pm$0.1 arcsec (measured peak to peak).  When this
ellipsoid is de-projected the true axial ratio of the shell is found
to be 1.26$\pm$0.08.

\subsubsection{Spectroscopy}
\begin{figure*}
\vspace {11.0truecm}
\includegraphics{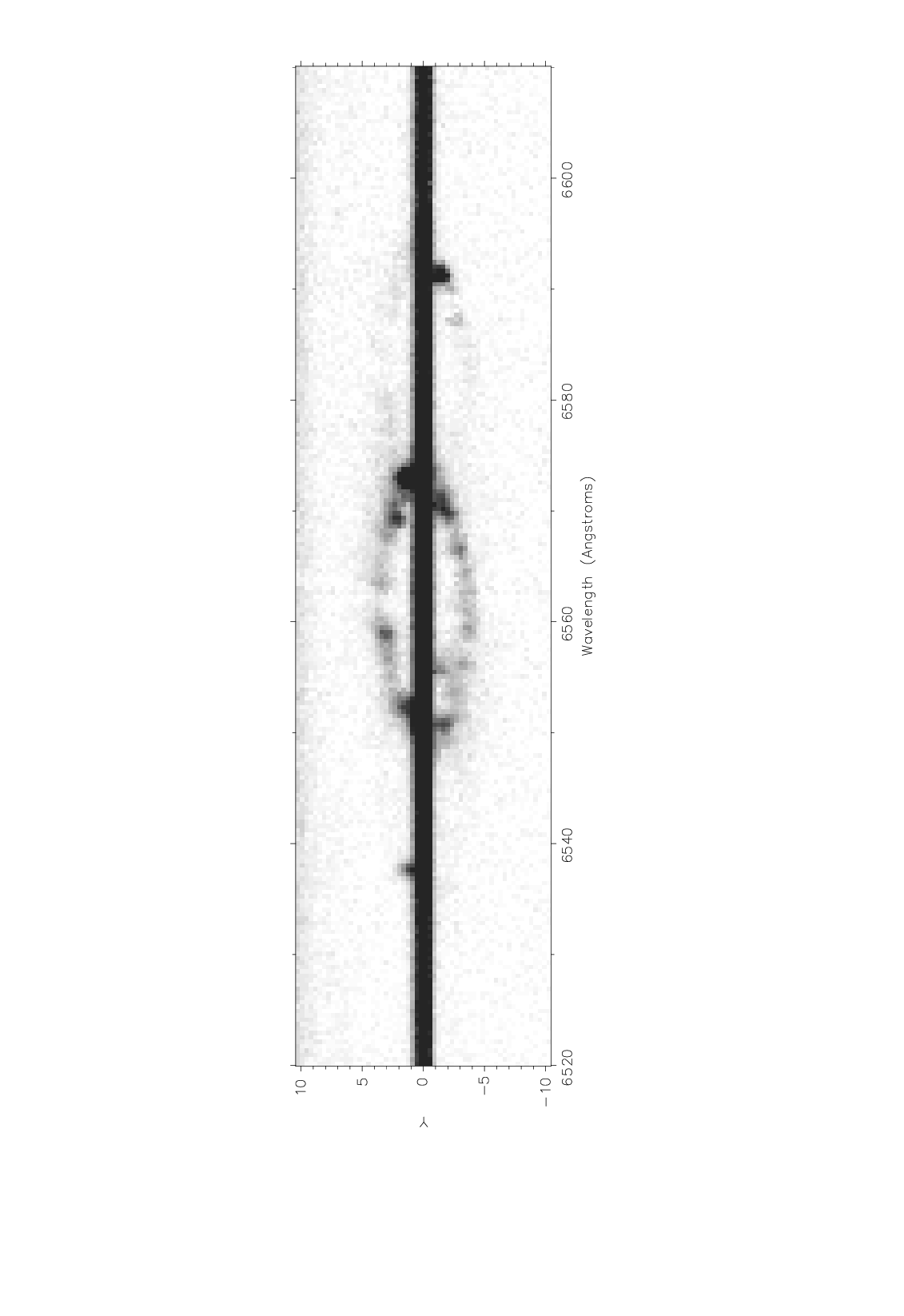}
\includegraphics{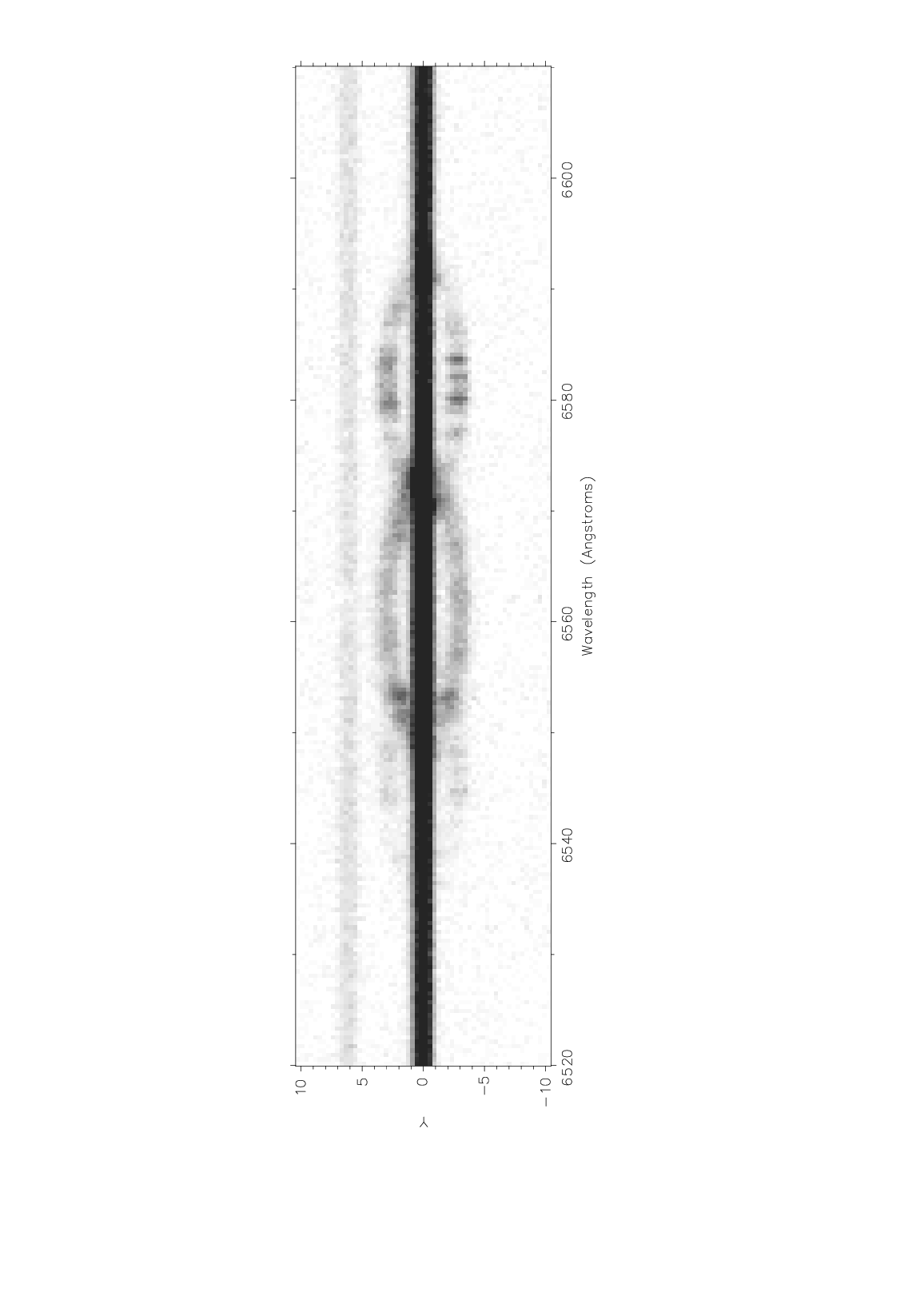}
\caption[]{Long slit spectra for the shell of FH Ser in slit positions 1
(top) and 2 (bottom). In slit 1 west is to the top of the frame and in
slit 2 north is to the top. The spatial direction is labelled as Y
in both slits and is marked in arcseconds.}
\label{fh_spec}
\end {figure*}

The long-slit spectra for both slit positions for the region of
6520--6610\AA~ are shown in Fig.~\ref{fh_spec}. 
Both slits show velocity ellipses of the shell in the lines of
H$\alpha$ ($\lambda$ 6562.8\AA) and
[N~II] ($\lambda$ 6549\AA~and $\lambda$ 6583\AA). 
The only other detection
of emission was from H$\beta$ which has the same appearance as that of the
H$\alpha$ ring. The two slit positions have been displayed using the same
high and low levels to allow relative comparison of fluxes (it should be
noted that the run was not photometric but a first order comparison is
possible). 

In slit position 1 the velocity ellipse in H$\alpha$ of the shell can
clearly be seen between the two fainter [N~II] ellipses. The H$\alpha$
ellipse shows clumpiness on small scales super-imposed on a well
defined velocity ellipse.  The [N~II] ellipses are both barely
detected apart from a feature above (blue-shifted) and below (red-shifted)
the central continuum source
corresponding to the equatorial ring seen in the HST image.  For the
$\lambda$ 6549\AA~ line these features appear at
$\lambda$$\sim$6538\AA~ and 6555\AA~(within the H$\alpha$ ellipse),
and for the $\lambda$ 6583\AA~line they appear at
$\lambda$$\sim$6573\AA~(co-incident with the red-shifted H$\alpha$
ellipse) and 6591\AA. The fact that the equatorial ring feature above
the stellar continuum (the west) is blue-shifted and the feature below
is red-shifted indicates that the shell is inclined so that the western
part of the shell is tilted away from the observer.

In slit position 2 the velocity ellipses are quite different to those
in slit 1. The northern and southern halves look the same and the
brightness of the [N~II] $\lambda$ 6583\AA~ellipse is comparable to
that from the H$\alpha$ ellipse. The H$\alpha$ ellipse seems to be of
fairly uniform intensity whereas the [N~II] ellipse is brighter near
the extremities of the shell (top and bottom of the slit). By
inspection of the image in Fig.\,\ref{fh_image} it can be seen that
the edges of the shell correspond with the region where the equatorial
ring is fully within the slit. The brightest region in the [N~II]
$\lambda$ 6583\AA~ellipse is brighter than any region in the H$\alpha$
ellipse.

\subsection{V533 Her}
\begin{figure*}
\vspace {8.0truecm}
\includegraphics{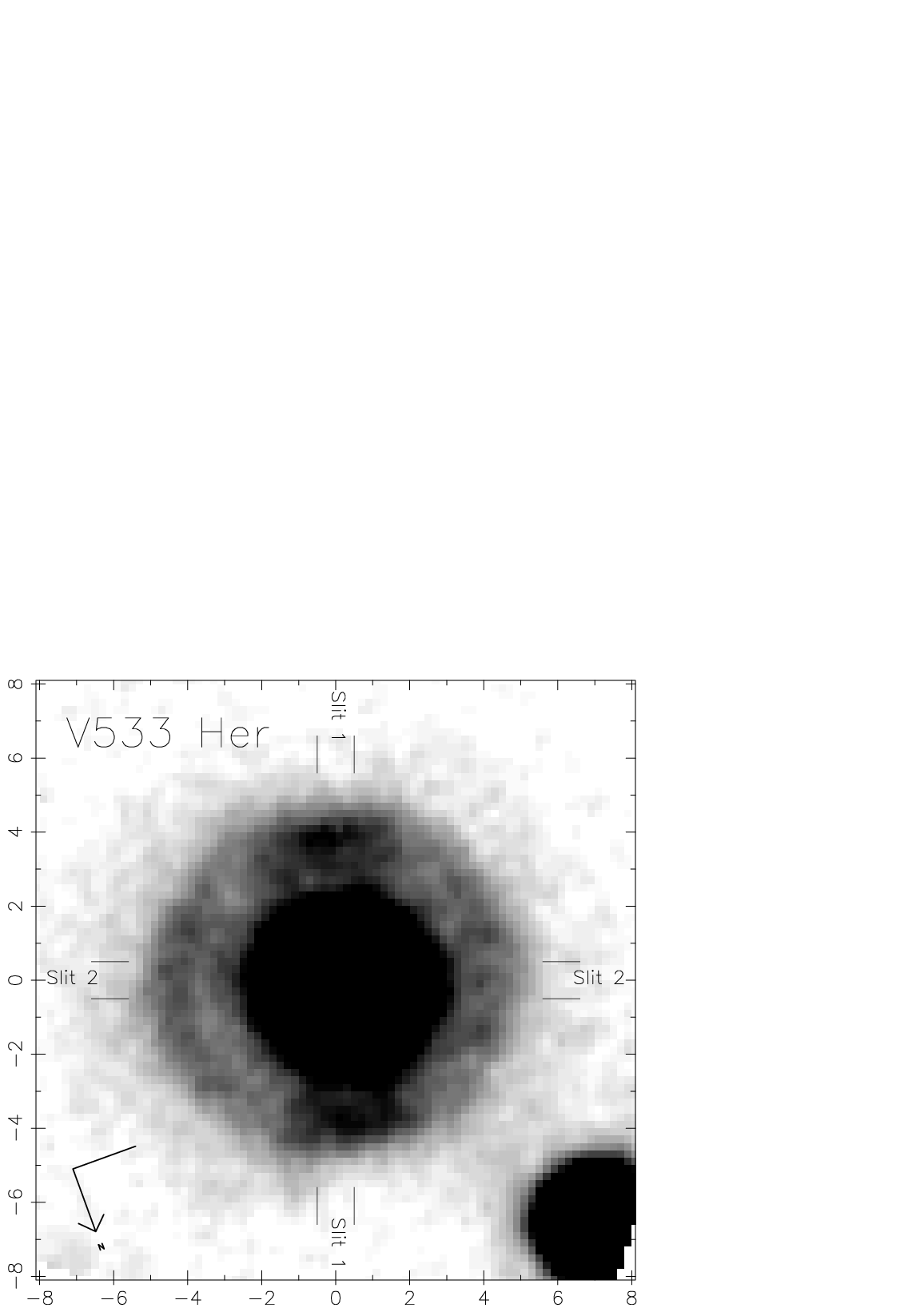}
\includegraphics{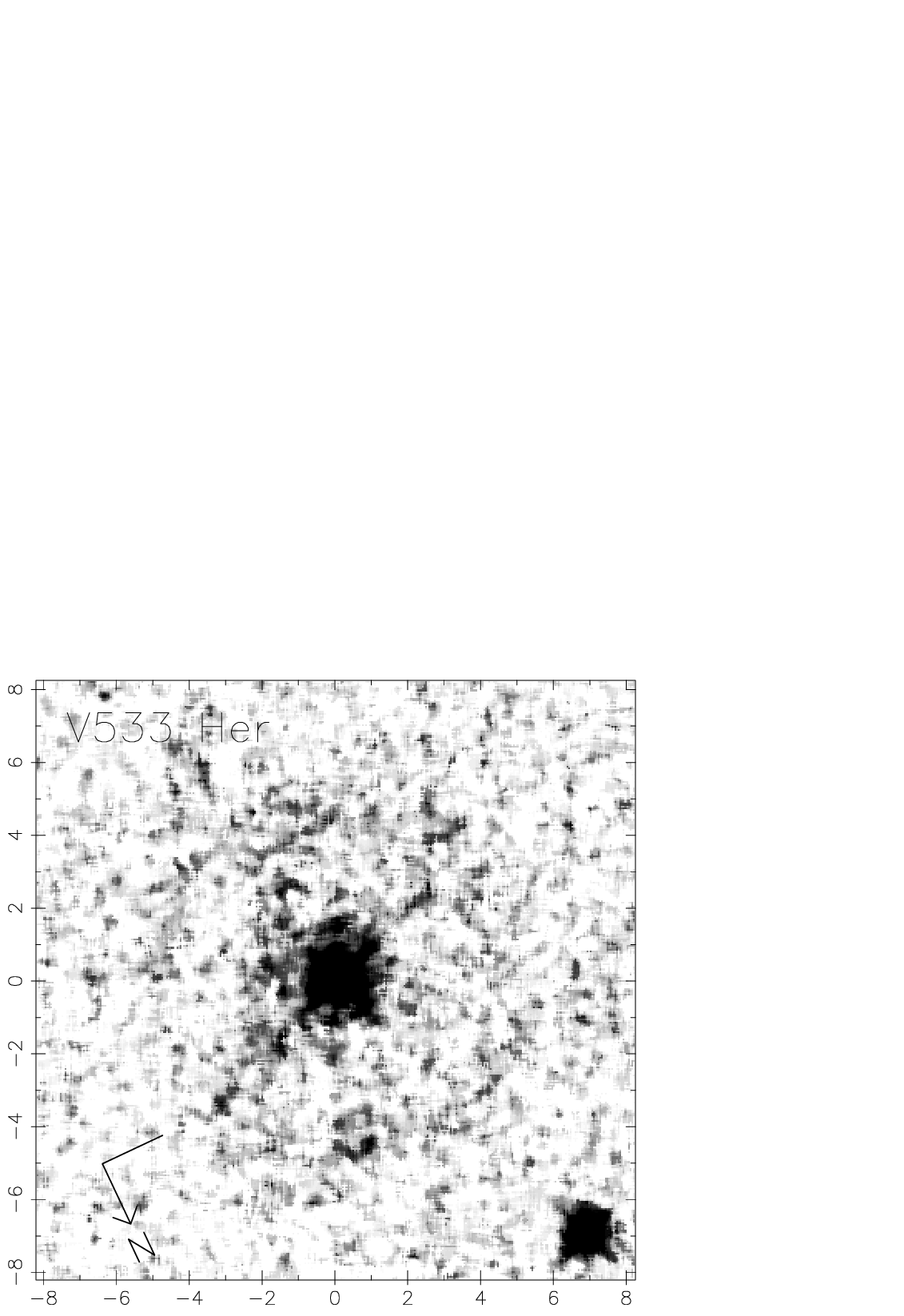}
\caption[]
{Ground-based image showing the positions of slits for the spectroscopy
of the V533 Her shell (left) shown next to HST F656N image of the V533
Her shell (right).  The ground-based image shown is a H$\alpha$
narrow-band image taken from Slavin et al.\ (1995). The HST image has
been processed with a 7$\times$7 sliding-box median-filter and is shown
scaled between the background and 5$\sigma$ above
the background. All axes are marked in arcseconds from the central
source. 
}
\label{v533_image}
\end {figure*}

V533 Her (Nova Herculis 1963) was discovered by Dahlgren (1963) and
extensively studied by Chincarini \& Rosino (1964).  It declined with
a $t_3$ time of 44d (see Duerbeck 1981 using AAVSO light curves). 
However combining
the V light curve presented by Chincarini (1964) which shows that it
declined to 6th magnitude sometime between March 4 and 8 with the
post-discovery results (see Haddock et al. 1963) that it actually
reached magnitude 3 on Jan 30 shows it may be more accurate to quote a
$t_3$ time of around 35d.  It has been classified as a `slow' nova
belonging to the {\it Fe II} spectroscopic class (Della Valle \& Livio
1997). The nebulosity around V533 Her was first detected by Cohen \&
Rosenthal (1983) by comparison with a point spread function. The first
clear image was provided by Slavin at al.\,(1995) who detected a
smooth circular shell, radius $\sim$4.5 arcsec, with a thick
equatorial band. This image is reproduced at the left of
Figure~\ref{v533_image}.

\subsubsection{Imaging}
V533 Her was imaged with HST for a total of 2600s with 2 cosmic ray split
observations on 1997 September 3. The median smoothed image of V533 Her
is shown at the right of Figure~\ref{v533_image}. An approximately
circular shell has been detected with radius $\sim$5 arcsec, consistent
with unifirm expansion since the Slavin et al.\ epoch. The
brightest knot at the bottom of the median-smoothed image of the shell
has a maximum value of 1.8 counts per pixel corresponding to a 7$\sigma$ 
detection at a surface brightness of 1.3$\times
10^{-15}$\,erg\,cm$^{-2}$\,s$^{-1}$\,arcsec$^{-2}$.
The signal to
noise in the HST image is however not sufficient to provide any more
information than was already apparent from the earlier ground-based
image.

\subsubsection{Spectroscopy}
WHT ISIS spectra were obtained on 1996 Aug 3. 
Two slit positions were used; slit 1 at PA
160$^\circ$ along the equatorial ring and slit 2 at PA 70$^\circ$
orthogonal to the equatorial ring. Both used  a slit
of width 1~arcsec. The two positions are 
indicated on the WHT image in Figure\,\ref{v533_image} reproduced from
Slavin et al. (1995) but rotated to correspond to the orientation of the
HST image shown alongside. 

\begin{figure*}
\vspace {7.0truecm}
\includegraphics{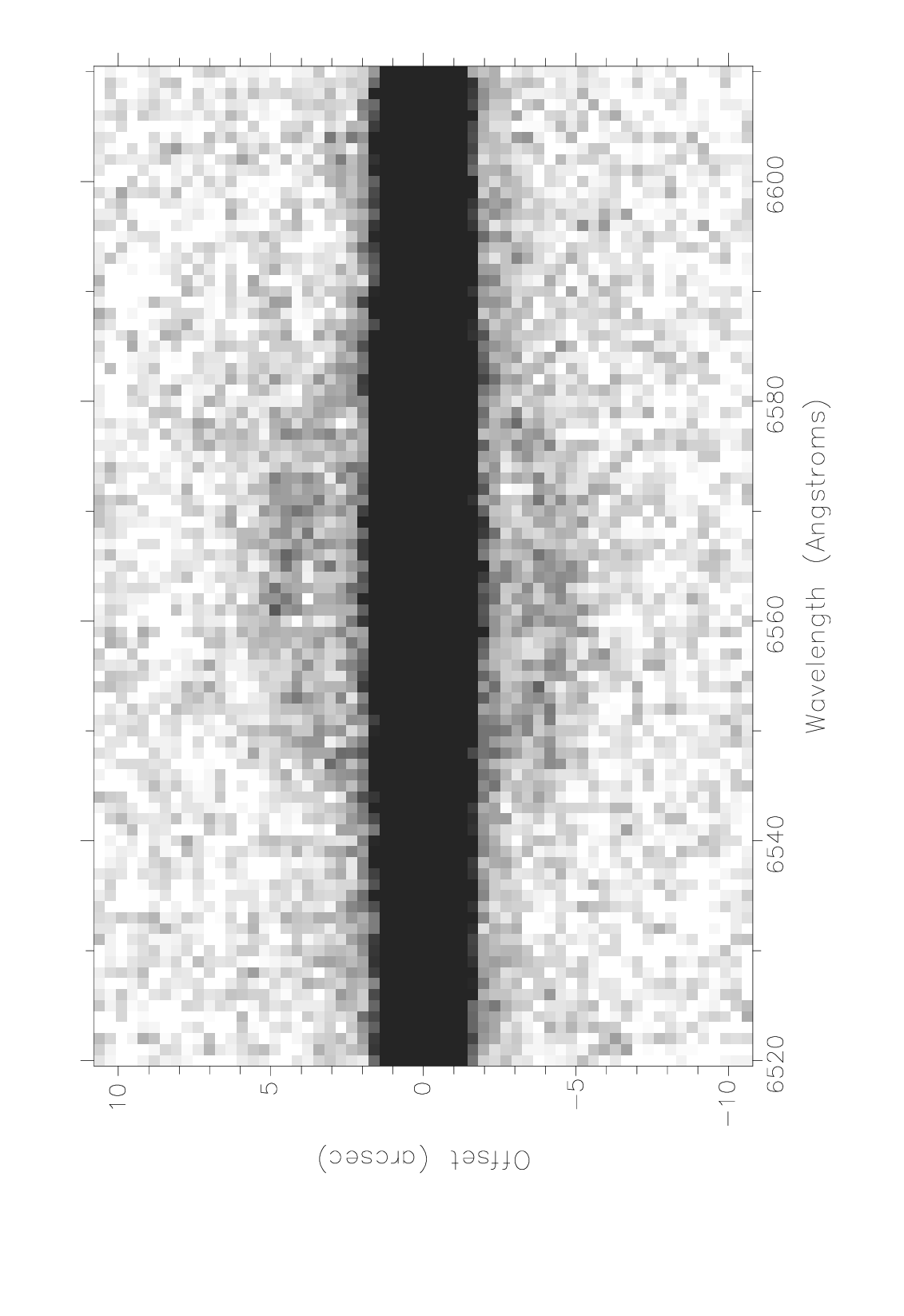}
\includegraphics{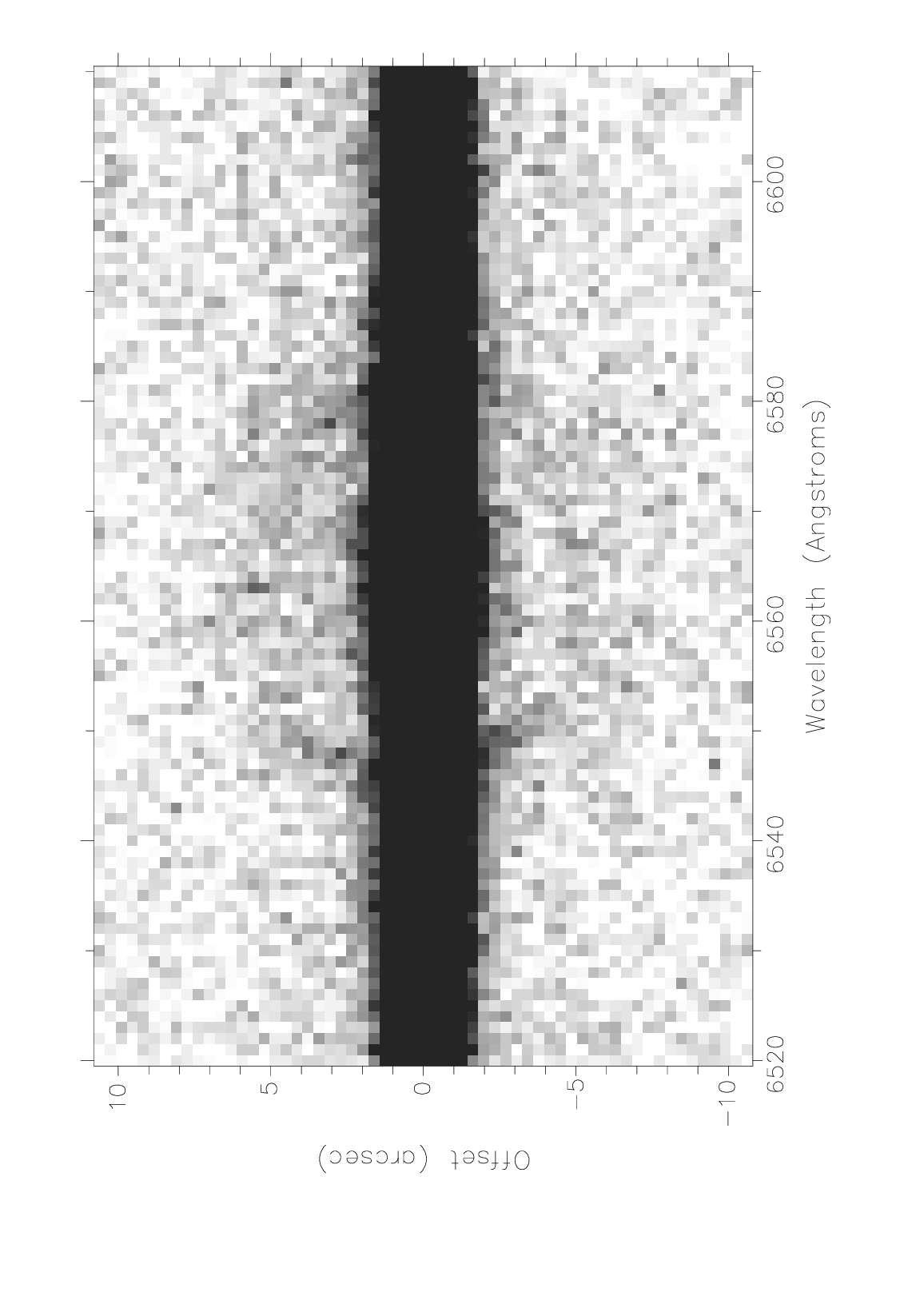}
\caption[Long slit spectra of V533 Her]
{Long slit spectra of the shell around V533 Her. Slit 1 (left) is
positioned across
the equatorial ring whereas Slit 2 (right) is perpendicular to the
equatorial ring. These spectra have been binned to 1\AA/pixel to make the
velocity ellipses more apparent.}
\label{v533_spec}
\end {figure*}

The spectra in the region of H$\alpha$ are shown in
Figure~\ref{v533_spec} (the H$\beta$ range is not reproduced here as,
although the shell is just visible, the signal to noise is very poor).
From these spectra we see that the radius of the shell at the first slit
position is $4.5\pm0.5$~arcsec whilst at the second slit position this
increases to $5.5\pm0.5$~arcsec. Hence there is some evidence that the shell
is elliptical with an apparent axial ratio of $1.2\pm0.2$. The major
axis is orthogonal to the equatorial band which is consistent with a
prolate ellipsoidal shell as observed in other novae e.g.\ FH Ser. 
We are unable to
estimate an inclination and hence a true axial ratio as the equatorial
band is not clearly detected in the HST image. The maximum line of sight
velocity (obtained from the points where the slits cross at the centre of
the nebula) is estimated to be $850\pm150$~km~s$^{-1}$. Expansion
parallax, taking into account the estimated errors in the shell diameter
and expansion velocity, leads to a distance estimate of $1250\pm300$~pc.

\subsection{BT Mon}
BT Mon reached maximum in 1939 September. The $t_3$ time is disputed in
the literature due to the exact date of maximum light being missed.
Payne-Gaposchkin (1964) quotes a $t_3$ of 36 days calculated from
spectral data whereas Schaefer \& Patterson (1983) quote a $t_3$ of 190
days from the light curve after maximum. The shell has been detected
around this nova in spectroscopic observations by Marsh et al.\ (1983).
The only image of the shell is that by Gill \& O'Brien (1998). 
They claimed detection of a ring of emission of diameter $\sim$7 arcsec
after deconvolution of the image with various techniques.

\begin{figure*}
\vspace {22.0truecm}
\includegraphics{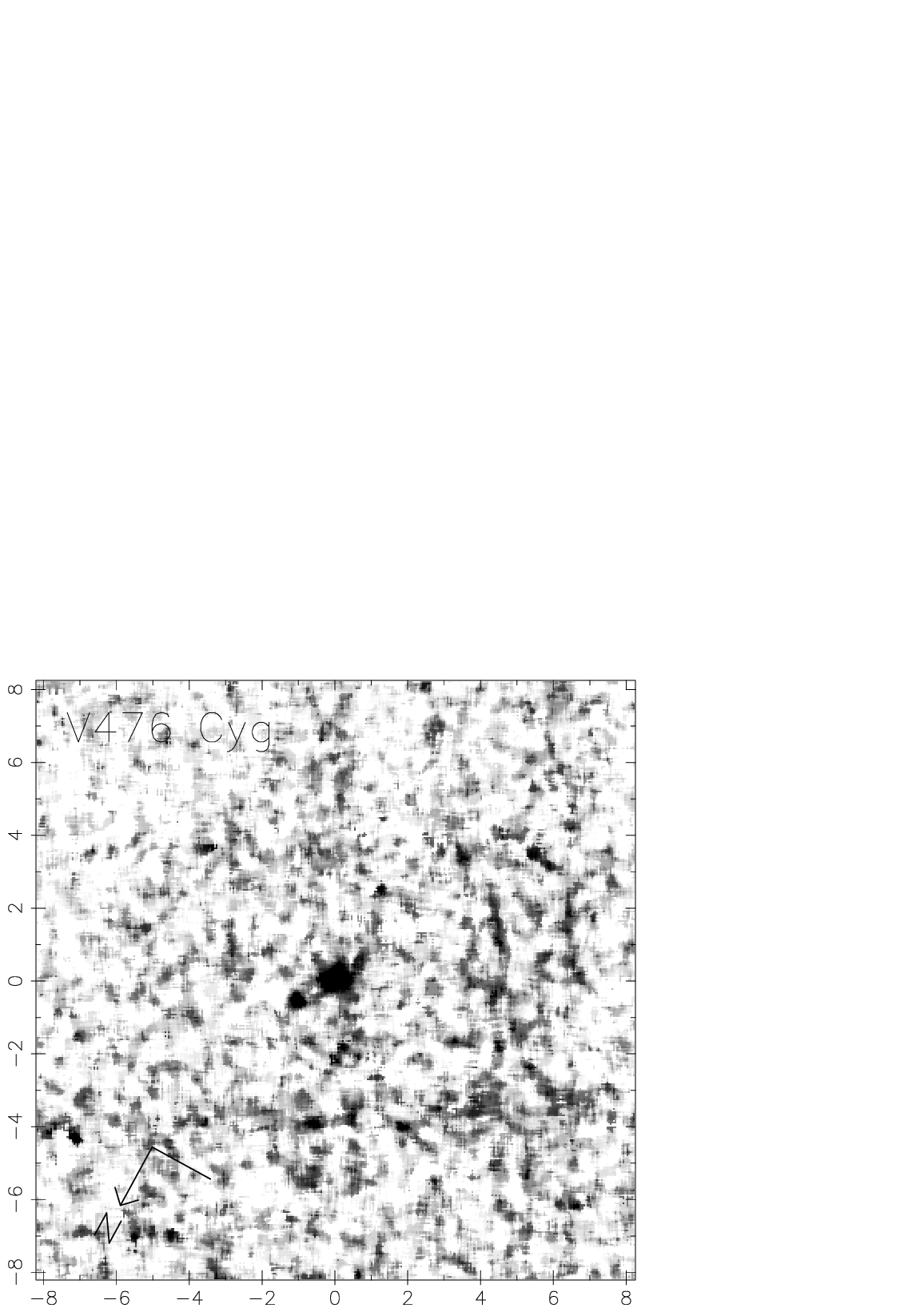}
\includegraphics{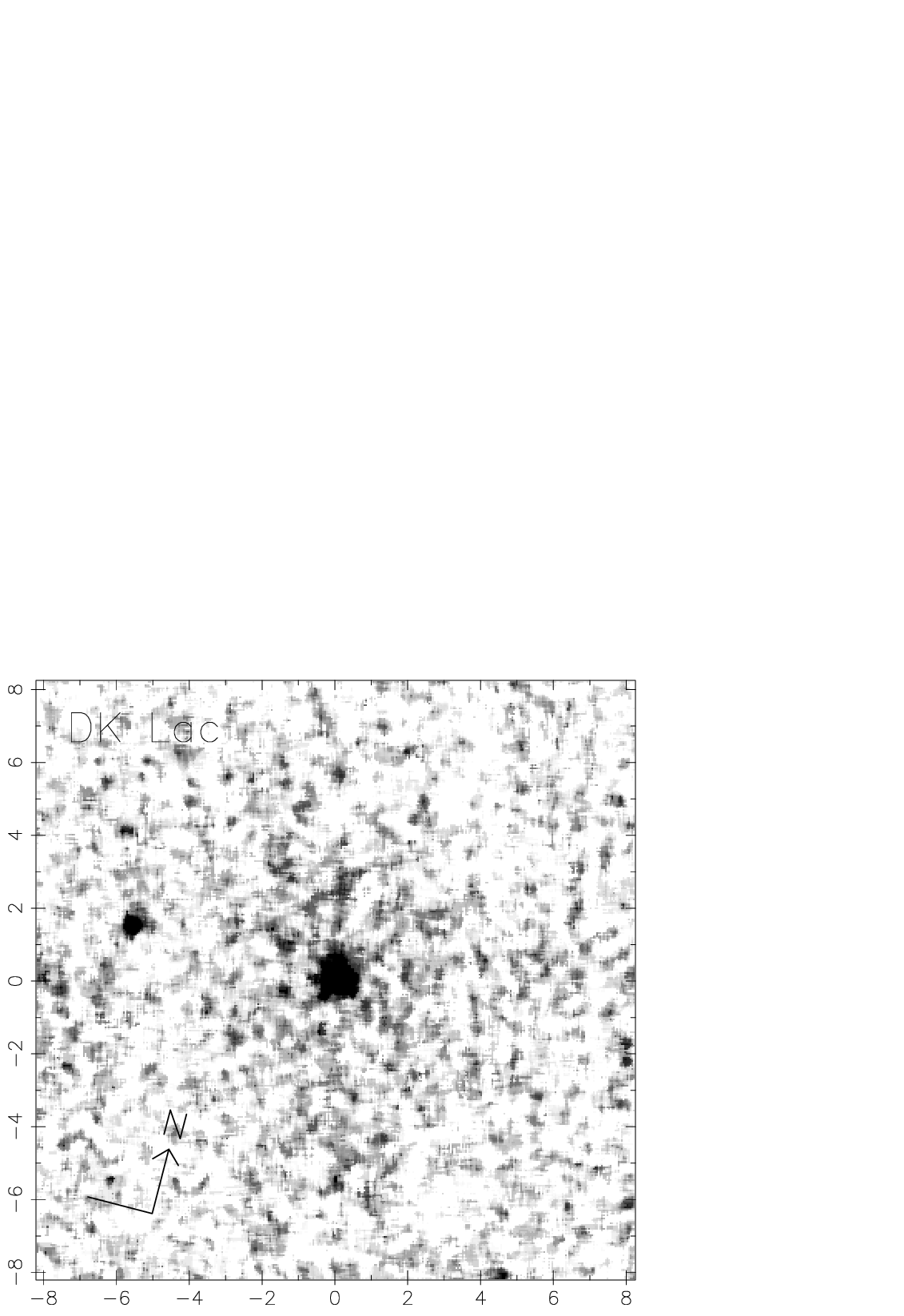}
\includegraphics{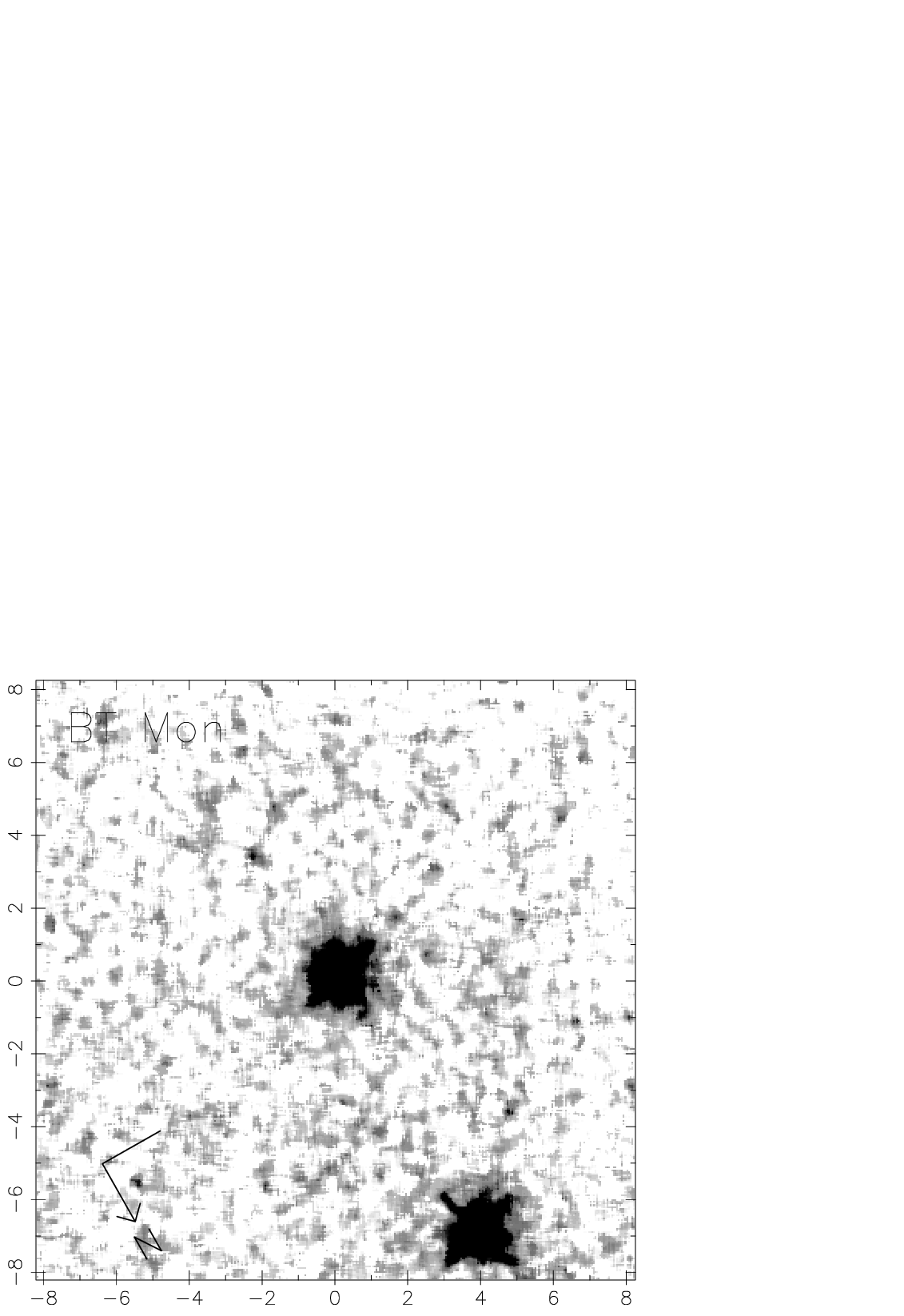}
\caption[]{
7$\times$7 sliding-box median-filtered F656N (H$\alpha$+[N II]) images
for V476 Cyg (top), DK Lac (middle) and BT Mon (bottom). All images are
scaled between the background level and 5$\sigma$ of the background above
the background level. All axes are marked in arcseconds from the central
source.}
\label{nondetect}
\end {figure*}

BT Mon was imaged with HST in two observations for a total of 2500s. The
median-smoothed image of BT Mon is shown in
Fig.\,\ref{nondetect}. No shell is detected putting a 3$\sigma$ upper 
limit on the surface brightness of the shell at 
6$\times 10^{-16}$\,erg\,cm$^{-2}$\,s$^{-1}$\,arcsec$^{-2}$.

\subsection{V476 Cyg}
V476 Cyg (Nova Cygni 1920) reached maximum on 1920 August 24. The
measured $t_3$ time for its decline was 16.5d (Duerbeck 1987)
classifying it as a very fast nova. A diffuse shell of radius $\sim 5$~arcsec
was detected by Slavin et al.\ (1995). With their
integration time of 900s it was barely detected above the 3 sigma level.

V476 Cyg was imaged with the HST for a total of 2600s (cosmic ray split by a
factor 2). The median-smoothed image of V476 Cyg is shown at the top left
of Fig.\,\ref{nondetect} where no nova shell is detected. This provides 
a 3$\sigma$ upper limit to the shell surface brightness of 
5$\times$10$^{-16}$\,erg\,cm$^{-2}$\,s$^{-1}$\,arcsec$^{-2}$.

\subsection{DK Lac}
DK Lac (Nova Lacertae 1950) was discovered during its rise in 1950 January.
The nova declined at a moderately fast rate with a $t_3$ of 32d,
accompanied by large variations of brightness. Slavin et al.\ (1995) 
detected the shell
around DK Lac, measuring is radius as 2.0 to 2.5 arcsec,
although the shell is poorly distinguished from the central source.

The median-smoothed HST image of DK Lac (total integration time 2500s) is
shown at the bottom left of Fig.\,\ref{nondetect}. No shell is detected
providing a 3$\sigma$ upper limit on the shell surface brightness of 
6$\times 10^{-16}$\,erg\,cm$^{-2}$\,s$^{-1}$\,arcsec$^{-2}$.

\section{Further Interpretation of the FH Ser Results}
\label{int}
\subsection{Observations} 
The [N~II] features seen in the spectrum from the first slit position (S1)
directly above and below the central source and the [N~II] features at
the extremities of the shell in the spectrum from the second slit position (S2)
are associated
with the equatorial ring clearly seen in the image. 
We would expect H$\alpha$ to mirror this
behaviour if the ring results from an enhancement in density.
However it can clearly be seen in
S2 that H$\alpha$ is not greatly enhanced in the ring at all. The
[N~II] ring features in S1 are approximately 40 times greater than the
typical [N~II] flux in the rest of the shell; 
although the H$\alpha$ ellipse is
rather clumpy in S1, there is no obvious
comparative increase in brightness in the corresponding regions for the
ring. Therefore the equatorial ring is dominated by enhanced 
emission in [N~II].

This picture is supported by the images in 
Slavin et al.\ (1995). In their figure
3c they show an image of the FH Ser remnant taken with a filter central
wavelength $\lambda$ 6560\AA , FWHM 17\AA. From examining the long slit
spectra shown in this  paper 
we can see that this will include almost all light in
H$\alpha$ apart from the parts of the shell moving directly towards or
away from us. There should be no [N~II] contamination. This image shows
only a smooth ring of emission resembling a simple limb-brightened shell.
There is no evidence in this image for an equatorial enhancement. Slavin
et al.\ also show an image with filter central wavelength $\lambda$
6537\AA , FWHM $>$ 17\AA~(their figure 3a). From our spectra we can see
that this corresponds primarily to blue-shifted [N~II] $\lambda$ 6549\AA. This
image shows a half-crescent of emission with very little emission in the
rest of the shell which again implies the equatorial enhancement is an 
[N~II] feature. This crescent of emission is to the west of the central
source which agrees with our suggestion that the eastern pole is tilted
towards us.

\subsection{Synthetic Images and Spectra}

Using the code described in Gill \& O'Brien (1999) it is possible to
produce synthetic images and spectra for simple models of the FH Ser nova
shell. The code calculates emission as a function of density and
therefore separate models needed to be produced for the H$\alpha$ and [N~II]
emission lines. The lines can be scaled relative to one
another by arbitrary factors and then combined, taking into account the
rest wavelengths of the lines and the Doppler shifting of each element,
to produce a spectral cube for a given wavelength range.

To generate the synthetic spectra the spectral cube was then smoothed in
the spatial directions with the equivalent of a 0.5 arcsec FWHM Gaussian
to simulate the seeing of the ground-based observations. A `slit' was
then positioned on the cube at positions and widths to correspond with
the observations. Smoothing must be performed before the slit is
positioned onto the cube to allow light to be scattered into and out of
the slit. The possibility of correction for a systemic velocity for any
model was incorporated to allow a best fit to the data.

Synthetic images were generated by convolving the unsmoothed spectral
cubes with the system throughput of the HST with the F656N filter
(Leitherer 1995). This was approximated by a profile starting with zero
throughput at $\lambda$ 6545\AA, rising linearly to peak at
$\lambda$ 6554\AA, remaining constant to $\lambda$ 6576\AA, then falling
linearly back to zero at $\lambda$ 6582\AA. The images were then
generated by collapsing the filtered spectral cube in the spectral
direction.

\begin{figure*}
\vspace {10.0truecm}
\includegraphics{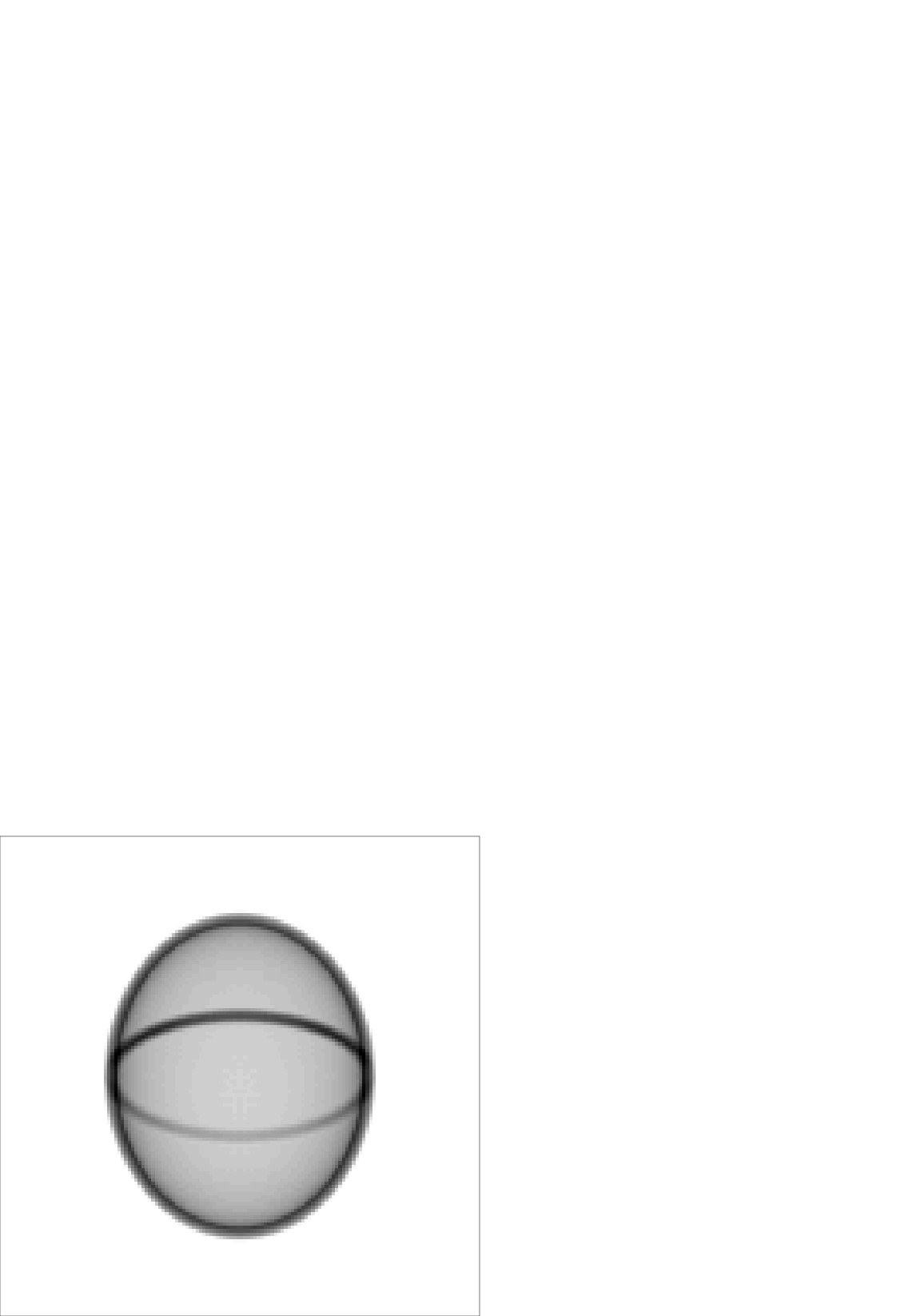}
\caption[]{Simulated HST F656N image of the remnant around FH Ser containing 
contributions from H$\alpha$ and [N~II]. }
\label{fh_im_syn}
\end {figure*}

\begin{figure*}
\vspace {11.0truecm}
\includegraphics{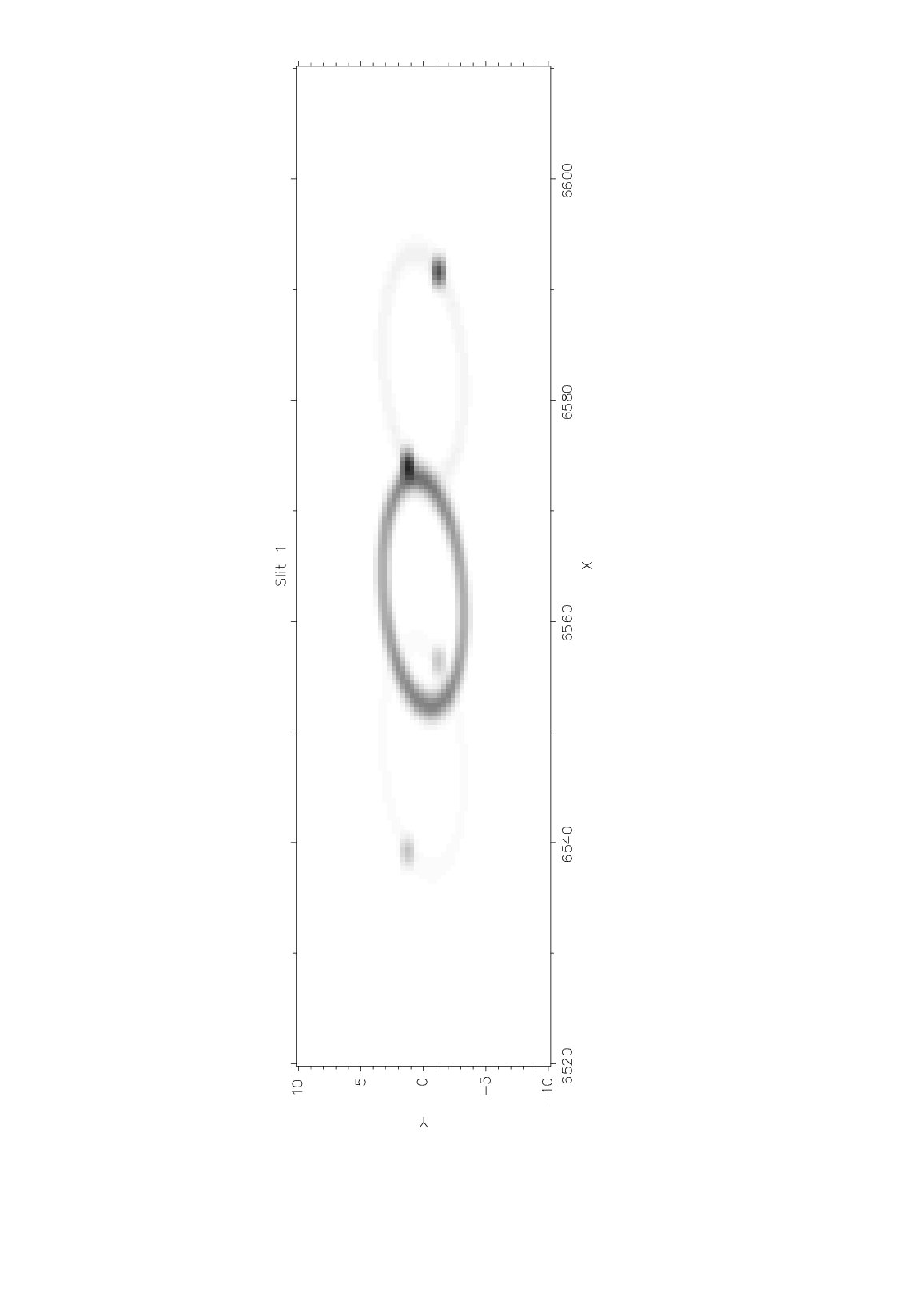}
\includegraphics{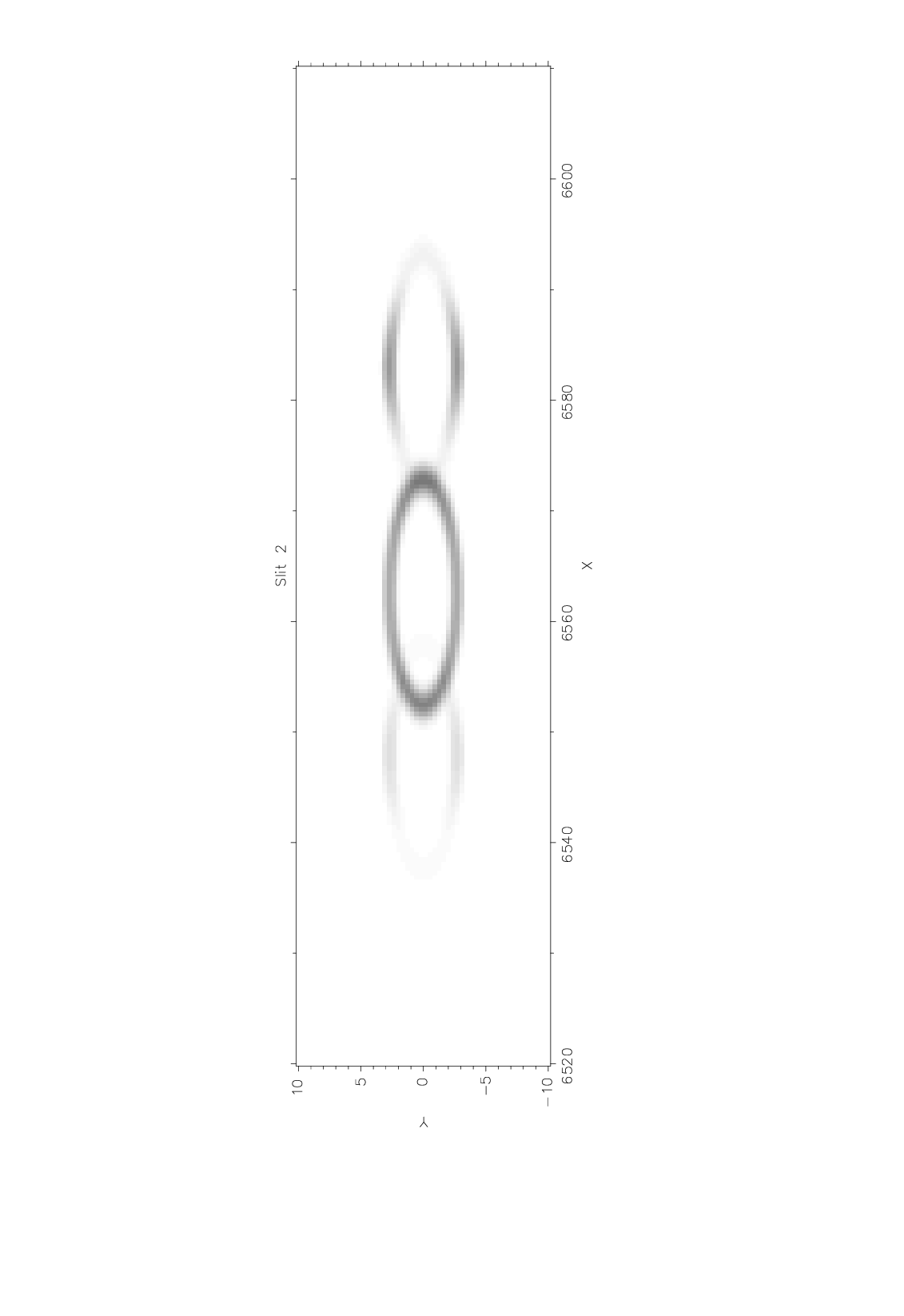}
\caption[]{Synthetic long slit spectra for the shell of FH Ser in slit
positions 1 (top) and 2 (bottom). In slit 1 west is to the top of the
frame and in slit 2 north is to the top. The spatial direction is labelled as
Y in both slits and is marked in arcseconds. The spectral
direction, labelled X, is marked in Angstroms. }
\label{fh_spec_syn}
\end {figure*}

We have generated models corresponding to a thin prolate ellipsoidal
shell with ellipticity 1.3 and inclination 62$^\circ$. For the
[N~II] components an equatorial enhancement by a factor of 40 in
intensity was also added in a thin ring. The light from H$\alpha$ was
increased by a factor of 8.6 relative to the typical (i.e.\ not the enhanced 
value appropriate to the ring) [N~II] $\lambda$ 6583\AA~
line (to best fit the observations), and the [N~II] $\lambda$ 6549\AA~
scaled by a factor 0.34 relative to the [N~II] $\lambda$ 6583\AA~ line.
Velocities were set proportional to the distance from the centre of the
ellipsoid and the maximum line of sight velocity was set to be 530 km/s.
For the model spectra to best fit the observations it was found that a
systemic velocity of $-$45 km/s was required. The resulting synthetic
spectra for positions S1 and S2 can be seen in 
Fig.~\ref{fh_spec_syn} and the synthetic
image in Fig.~\ref{fh_im_syn}. The model agrees closely with
the observations shown in Fig.\,\ref{fh_spec}. 
This also allows us to derive a distance to FH Ser of 
950$\pm$50 pc which takes full account of the asphericity of the shell.

If we examine the observed image in Fig.\,\ref{fh_image} 
we see that the equatorial ring is 
brighter in one half than the other, behaviour replicated in our synthetic 
image in Fig.\,\ref{fh_im_syn}. Our model shows that this is 
because the system
throughput of the HST with the F656N filter lets through the red-shifted
$\lambda$ 6549\AA~light and the blue-shifted $\lambda$ 6583\AA~light.
Therefore the ring seen in the image isn't actually a whole ring from one
line but two halves of the ring each from a different line.
The ratio of the lines in this doublet is always $[6549]/[6583]=0.34$, so
the half of the ring visible in the image which is 
produced by the $\lambda$ 6583\AA~line is 3 times
brighter than the $\lambda$ 6549\AA~half. 

\subsection{The origin of the [N~II] ring}

For the equatorial ring to be enhanced in [N~II] relative to the rest of
the shell requires either a density, ionization or metallicity effect. 

A simple density increase would produce more [N~II] (providing that the
level did not start to become collisionally de-excited) but also more
H$\alpha$ and is therefore presumably not the solution.

An increase in the amount of nitrogen in the equatorial plane of the FH
Ser shell could explain the [N~II] enhancement. There is the possibility of a
metallicity variation across the surface of the white dwarf before
eruption due to rotation of the white dwarf 
although it is difficult to see how a
thin enhancement of nitrogen could be produced rather than a more wide-spread 
variation around the shell.

The final possibility is an ionization effect which would require either
an increase or decrease in ionization in the equatorial region. A good
candidate for providing such an effect is the accretion disc around the
central source. An explanation could be that the main shell is very
highly ionized, doubly ionized or above, by the central parts of the
accretion disk. The equatorial region is then shadowed from this hot
central UV source, causing the relative abundance of 
nitrogen in the singly ionized state to be increased. 
This would then produce an
increase in [N~II] emission in a thin equatorial region with no increase
in H$\alpha$. This proposal needs testing by further observation and 
modelling. A similar suggestion was made by Petitjean et al.\ (1990) for 
the shell of DQ Her.

\section{Conclusions}
\label{conc}
HST WFPC2 images and WHT spectroscopy of FH Ser have allowed us to
derive a consistent model for its shell as an ellipsoid with
ellipticity 1.3$\pm$0.1, and inclination 62$\pm$4$^\circ$. The
equatorial expansion rate is found to be 490$\pm$20~km~s$^{-1}$ and
the systemic velocity $-45$~km~s$^{-1}$. This gives a distance to FH
Ser of 950$\pm$50~pc. These values agree within the
errors with the results presented in previous papers (Della Valle et
al.  1997, Slavin et al. 1995, Duerbeck 1992).  The equatorial ring
seen in the HST images and previous ground-based images is found to be
due to a 40$\times$ increase in emission from [N~II] rather than
H$\alpha$+[N~II]. It is suggested that the origin of this ring lies
either in abundance gradients in the shell or in aspherical
illumination of the shell by the central photoionizing source. This
latter suggestion will be developed in a paper to follow.

Similar imaging and spectroscopy of the nova V533 Her reveal a shell of 
radius $\sim$5~arcsec with an axial ratio of $1.2\pm0.2$. The expansion
velocity of this shell is $850\pm150$~km~s$^{-1}$ and the distance is estimated
to be $1250\pm300$~pc. The quality of the data is however not sufficient for 
us to examine the structure of the shell in as much detail as was possible 
for the shell of FH Ser.  

Our HST images of BT Mon, V476 Cyg and DK Lac provided only upper
limits on the surface brightness of their shells. This may seem
surprising given that these shells had been previously detected from
the ground (Gill \& O'Brien 1998, Slavin et al. 1995). However it
should be noted that the ground-based images were taken using the large 
aperture 4.2m WHT and 3.9m Anglo-Australian Telescope.  Although the improved
spatial resolution of HST helps overcome its reduced collecting area when
the nebula is clumpy and these clumps are unresolved from the ground
this does not apply when the emission is more smoothly distributed. These 
nova shells were not particularly well-detected in the ground-based survey
(BT Mon requiring image deconvolution) so it is perhaps not too 
surprising that we did not detect them here.

Other novae to be investigated in the following papers of this series include
HR~Del, RR~Pic, T~Aur, V842~Cen and V1500~Cyg. 

\section {Acknowledgements}
We would like to thank Dr M. Della Valle for valuable comments at
the refereeing stage. CDG would like to thank PPARC for his
studentship, and Nial Tanvir and Rachel Johnson for their help with
the reduction of the HST images.  This work is based on observations
with the NASA/ESA Hubble Space Telescope (obtained at the Space
Telescope Science Institute, which is operated by the Association of
Universities for Research in Astronomy, Inc. under NASA contract
No. NAS5-26555) and the William Herschel Telescope (operated on the
island of La Palma by the Isaac Newton Group in the Spanish
Observatorio del Roque de Los Muchachos of the Instituto de
Astrofisica de Canarias).

\newpage

\begin{table*}
\begin{tabular}{llll}
\multicolumn{4}{c} {\bf HST WFPC2 Imaging }\\
{\bf Nova}	&{\bf Epoch } & {\bf Filter} & {\bf Integration time} \\[0.1cm]
FH Ser & 1997 May 11 & F656N & 2$\times$1200s\\
V533 Her & 1997 Sep 3 & F656N & 2$\times$1300s\\
BT Mon & 1997 Apr 26 & F656N & 2$\times$ 1200s\\
DK Lac & 1998 Mar 28 & F656N & 2$\times$800s $+$ 1$\times$900s\\
V476 Cyg & 1997 Aug 5 & F656N & 2$\times$1300s\\[0.1cm]
\end{tabular}
\vspace*{0.3cm}
\begin{tabular}{llllll}
\multicolumn{6}{c} {\bf WHT ISIS Spectroscopy }\\
{\bf Nova}	&{\bf Epoch } & {\bf Wavelength range} & {\bf Grating} & 
{\bf Integration time} & {\bf Slit PA}\\[0.1cm]
FH Ser & 1996 Aug 3 & 4740--5160\AA & 1200B & 2$\times$900s & 84$^\circ$ \\
       &             & & &               & 174$^\circ$ \\
       &             & 6390--6810\AA & 1200R & 2$\times$900s & 84$^\circ$                \\
       &             & & &               & 174$^\circ$ \\
V533 Her & 1996 Aug 3 & 4740--5160\AA & 1200B & 2$\times$1800s & 160$^\circ$\\[0.1cm]
         &            &               &       &                & 70$^\circ$\\[0.1cm]
         &            & 6390--6810\AA & 1200R & 2$\times$1800s &  160$^\circ$\\[0.1cm]
         &            &               &       &                &  70$^\circ$\\[0.1cm]

\end{tabular}
\caption{Details of the HST WFPC2 imaging and WHT ISIS spectroscopy presented 
in this paper. The F656N filter has central wavelength 6560~${\rm\AA}$ and 
full width at half maximum of 30~${\rm\AA}$. Slit widths were 1 arcsecond.}
\end{table*}

\newpage
\newpage

\end{document}